\documentclass[prl,twocolumn,graphicx]{revtex4}

\usepackage{amsmath}
\usepackage{enumerate}
\usepackage{amssymb}
\usepackage{graphicx}
\usepackage{wrapfig}
\usepackage{setspace}
\usepackage[hypertex]{hyperref}

\renewcommand{\>}{\rangle}

\begin{document}

\title{Multi-condensate states in BCS superconductors}

\author{E. Bettelheim}
\affiliation{Department of Physics, Hebrew University,
Jeruslaem, Israel}

\date{\today}

\begin{abstract}

A BCS (Bardeen-Cooper-Schrieffer) superconductor, which is placed out of equilibrium, can develop quantum instabilities, which manifest themselves in  oscillations of the superconductor's  order parameter (pairing amplitude $\Delta$). These instabilities are manifestations of the Cooper instability. Inelastic collisions are essential in resolving those instabilities. Incorporating the quantum instabilities and collisions in a unified approach based on Richardson's exact solution of the pairing Hamiltonian, we find that a BCS superconductor may end up in a state in which the spectrum has more than one gap.

\end{abstract}

\maketitle

\section{Introduction}
An equilibrium BCS superconductor has the property that the spectral gap is located around the Fermi energy. When the superconductor is connected to leads thorough a tunnel junction and current is allowed to pass through the system, the super-conducting gap may be found below or above the Fermi energy. Since excitations above the gap are electron-like and those below the gap are hole-like, the shift of the gap away from the Fermi energy is accompanied by a net difference, $\Delta N$, between the number of hole-like and electron-like excitations. A quantity 'branch imbalance' with the dimension of energy is denoted by $\Phi$ and defined by $\Phi \equiv \frac{\Delta N}{\rho_0}$, where $\rho_0$ is the density of states.

Experiments \cite{Clarke, ChiClake} have directly demonstrated the difference between the Fermi energy and the energy of the gap (or the condensate energy) in the same piece of super-conducting material. In order to understand how the injection of excitations through a tunnel junction into a superconductor can lead to imbalance, consider first that the current is injected into the superconductor as normal current while in the superconductor it flows as super-current.   This means that a process by which normal excitations are converted into condensed pairs must take place in the injection region. The condensed pairs then flow away from the injection region as super-current.

The processes responsible for converting normal excitations into condensed pairs are collisions, in many case predominately electron-phonon collisions. These collisions constantly convert electron-like excitations into condensed pairs or destroy condensed pairs by converting hole-like excitations. The energy of the condensate, and consequently of the condensed pairs, can be assigned as the energy of the gap (the gap midpoint). If the gap resides at the Fermi energy, particle-hole symmetry leads to equal rates for condensed-pair creation and annihilation, i.e. to no net conversion of quasi-particles into condensed pairs. It is the shift of the gap relative to the Fermi surface that allows for a net conversionof quasi-particles into condensed-pairs.

Close to the critical temperature and under certain assumptions, to be detailed below, one can write down an approximate distribution function for the excitations describing a situation where the quasi-particles are at equilibrium at chemical potential $\xi_1-\Phi$, while the condensate is at energy $\xi_1$ \cite{galaiko} \cite{spivak} \cite{tinkham}:
\begin{align}\label{OneGapDistr}
n(\xi) = \frac{1}{1+e^{\frac{\epsilon(\xi)+\Phi q(\xi)}{kT}}},
\end{align}
where $\epsilon(\xi)$ and $q(\xi)$ are the energy and charge of an excitation, respectively:
\begin{align}\label{spectrumandcharge}
\epsilon(\xi) = \sqrt{(\xi-\xi_1)^2+\Delta_1^2}, \quad q(\xi) = \frac{\xi-\xi_1}{\epsilon(\xi)}.
\end{align}
$\Delta_1$ denotes the size of the gap. Quasi-neutrality demands $\Phi=\xi_1$.

\begin{figure}
\includegraphics[width=2.0in]{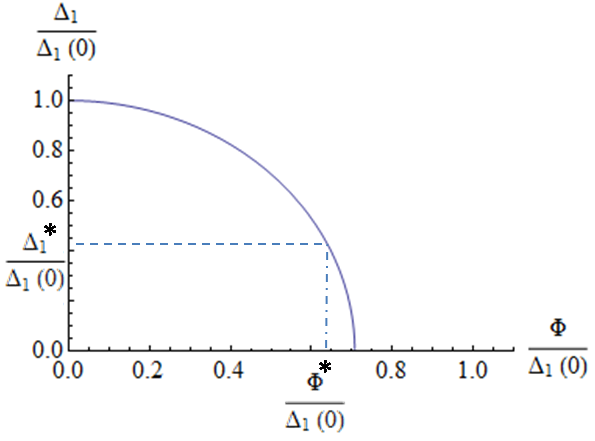}
\caption{Dependence of gap, $\Delta_1$, on imbalance, $\Phi$, in units of the gap in the absence of imbalance, $\Delta_1(0)$. Beyond the critical value of the imbalance, $\Phi^*$, the systems is not stable for Cooper-pairing at the Fermi energy.    }
\label{SemiCircle}
\end{figure}

Eq. (\ref{OneGapDistr}) is only an approximation and has corrections, but for a qualitative understanding of the effect we want to describe, it shall be sufficient to ignore these for the time being. We follow here closely the treatment of \cite{spivak}. If one plugs (\ref{OneGapDistr}) into the self-consistency equation, one obtains a relation between $\Delta_1$ and $\Phi$, as follows:
\begin{align}\label{simplesupression}
\Delta_1^2=\Delta_1(0)^2-2\Phi^2,
\end{align}
where $\Delta_1(0)$ is the gap at $\Phi=0$. It is seen that branch imbalance, as measured by $\Phi$, suppresses the gap (see Fig. (\ref{SemiCircle})). In fact at $\Phi=\Delta(0)/\sqrt{2}$, the gap is completely suppressed, and all pairing correlations vanish. The system returns back to the normal metal state. It is known, however, that this state is unstable. Any normal metal placed below $T_c$ is unstable to Cooper pairing, i.e. to formation of pairing correlation at the Fermi energy.  There must be a critical value of $\Phi$, designated as $\Phi^*$, above which, the system becomes unstable to such pairing, as pointed out in \cite{spivak}.

The scenario associated with the instability can be described roughly as follows: pairing correlations are restricted to a region of size comparable to the gap, $\Delta_1$ at the gap energy, $\xi_1$. As $\xi_1=\Phi$ becomes larger, this region becomes  smaller because of suppression. As a consequence the Fermi energy remains devoid of pairing correlations. The cooper instability -- an instability to the formation of correlations at the Fermi energy --  then takes place.

The purpose of this letter is to analyze the state which is formed beyond the instability point. We shall show,  that a novel state appears where another gap, $\Delta_2$, is formed at an energy $\xi_2$ closer to the Fermi level. We shall make use of Richardson's exact solution of the pairing Hamiltonian, to describe such a state. Other approaches are known to be inadequate to describe the state beyond the instability point \cite{aronov, PethickSmith}.

\section{Semiclassical approach for instability}
To understand better the instability that takes place, it is useful to consider the behavior of Anderson's pseudo-spins  \cite{anderson}:
\begin{align}
s_j^z =\frac{1}{2} \left( \langle \sum_\sigma c^\dagger_{j,\sigma}c_{j,\sigma} \rangle - 1 \right), \quad s_j^+ \equiv s^x_j -i s^y_j = \langle c^\dagger_{j,\downarrow} c^\dagger_{j,\uparrow} \rangle, \nonumber
\end{align}
which afford a semiclassical description. The order parameter $\Delta$ is given by $\Delta \equiv  \Delta_x -i \Delta_y \equiv \frac{g}{2} \sum_j s^+_j$, and is a measure of the overall pairing correlations.   The semiclassical  limit of the dynamics may be obtained by recasting the pairing Hamiltonian as follows:
\begin{align}\label{Hamiltonian}
H = \sum_{j, \sigma} \xi_j c^\dagger_{j,\sigma} c_{j,\sigma} -g \sum_{ j,j'} c^\dagger_ {j\uparrow} c^\dagger_{j\downarrow} c_{j'\downarrow}c_{j'\uparrow},
\end{align}
through the pseudo-spins, $H = \sum_{j=1}^{2M} 2\xi_j s^j_z -\frac{2}{g}|\Delta|^2$. This leads to the following equations of motion for the pseudo-spins:
\begin{align}\label{GorkovsEquaitons}
\dot{\vec{s}}_j =-2 \left(
\Delta_x,
\Delta_y ,
-\xi_j \right) \times \vec{s}_j
\end{align}
A  solution of  (\ref{GorkovsEquaitons}), where the order parameter takes the form $\Delta = \Delta_1 e^{2 i \xi t}$. Namely, the order parameter has a time-independent modulus, which equals the super-conducting gap, $\Delta_1$. This solution is provided by taking
\begin{align}\label{timeindependent}
\vec{s}(\xi) = \frac{n(\xi)}{\epsilon (\xi)} \left(\mbox{Re}(\Delta),\mbox{Im}(\Delta), \xi_1-\xi \right),
\end{align}
for $\Delta_1$ satisfying the self consistency condition:
\begin{align}
\frac{2}{g} = \sum_{\xi} \frac{n(\xi)}{\epsilon(\xi)}
\end{align}
Here $\epsilon(\xi)$ is given by (\ref{spectrumandcharge}), and $n(\xi)$ may be interpreted as the distribution function for the excitations.

The distribution function given by (\ref{OneGapDistr}) represents just such a solution (with time-independent modulus of the order parameter). The solution (\ref{timeindependent}) becomes unstable for $\Phi>\Phi^*$. The instability manifests itself in a linear stability analysis around the solution (\ref{timeindependent}). The unstable modes can be seen to represent pairing correlations forming at the Fermi energy. Namely such a mode contains non-vanishing components of $s^+(\xi)$ at energies around the Fermi energy. A perturbation around the time-independent solution will excite these modes. The pairing correlations around the Fermi surface then manifest themselves in a change of the order parameter $\Delta$. Moreover the modulus of the order parameter will cease to be time-independent.

\begin{figure}
\includegraphics[width=3.0in]{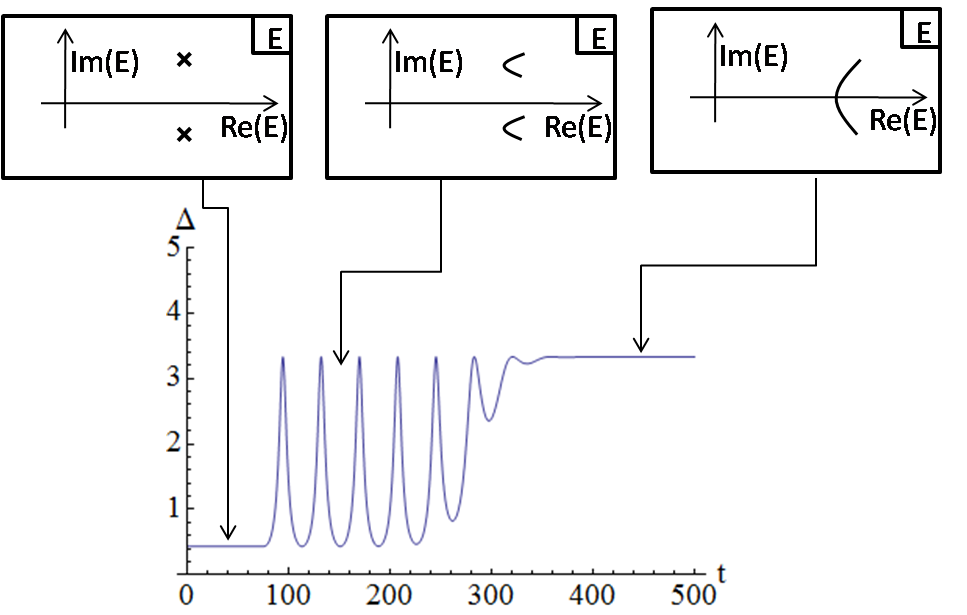}
\caption{A sketch of the time dependence of the modulus of the order parameter as a function of time, the typical Richardson spectrum is shown for each period. After an initial quiet period (left Richardson spectrum), a perturbation causes a train of solitons to appear (middle Richardson spectrum), these oscillations slowly die out and the global steady state ensues (right Richardson spectrum).  In the Richardson spectrum plots, $X$ denote instability points, while arcs are shown in contoured heavy lines. }
\label{Sketch}
\end{figure}

If instead of (\ref{OneGapDistr}) one takes the distribution function of the normal metal (obtained by taking $\Delta_1=0$, $\xi_1=\Phi=0$) one recovers the usual Cooper instability. Such an instability arises if one takes a normal metal and suddenly switches on the pairing interaction. This can be potentially realized in the lab by employing the Feshbach resonance \cite{ResonantControl, TunableInteraction}. Such a situation was considered in \cite{bls}. It was shown there that the time dependence of the order parameter exhibits an oscillatory behavior which can be described as soliton trains (See figure (\ref{Sketch})). It should be noted that this behavior is only exhibited at times before collisions take place. Collisions will tend to equilibrate the system bringing it to the BCS state, with its time-independent order parameter. To treat collisions one must go beyond (\ref{GorkovsEquaitons}), however in the simple Cooper instability problem one already has a good qualitative understanding of the system's behavior: First soliton trains will ensue, the oscillations associated with the soliton trains will slowly die out and the thermal equilibrium state, where the order parameter is constant in time, will be reached.

An important development in the study of the semiclassical equations is the discovery of their integrability. This was achieved in \cite{kuznetsov} and then utilized in  \cite{Kuznetsov2}. An interesting Wigner function formalism for the collisionless dynamics has appeared in \cite{KijkoWigner}.

In the case at hand, where the Cooper instability takes place at the same time that a gap already exists away from the Fermi surface, a qualitative understanding of the steady state is more complicated. The semiclassical analysis just after the instability takes place was studied in \cite{NahumandI}. However, without employing some new theory, no obvious candidate for the long time steady state beyond the instability point exists. Indeed (\ref{OneGapDistr}) does not provide such a solution for any value of $\Phi$ and $\Delta_1$. The purpose of the letter is to characterize this state. In particular we will find that two gaps coexist in the spectrum.

\section{Quantitative analysis of imbalance}
We shall treat the problem using a kinetic approach. In order for a kinetic approach to be valid, the time, $\tau_\Delta = \frac{\hbar}{\Delta}$, it takes for the spectrum to adjust to changes of the order parameter must be much smaller than the collision time $\tau_\epsilon$.
\begin{figure}
\includegraphics[width=2.4in]{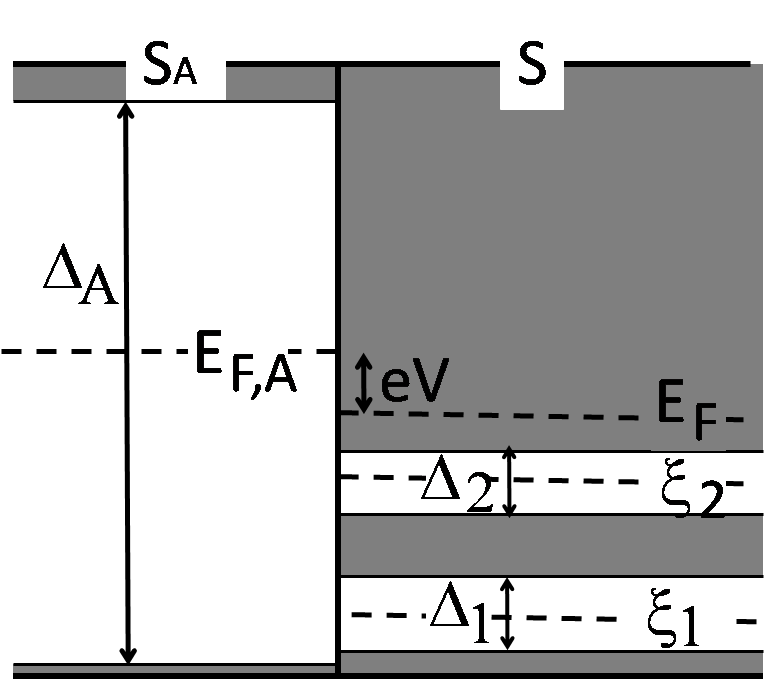}
\caption{
Two superconductors $S$ and $S_A$  with Fermi energies $E_F$ and $E_{F,A}$, are coupled through a tunnel junction. $S$ develops two gaps $\Delta_1$ and $\Delta_2$, at energies $\xi_1$ and $\xi_2$, respectively. $S_A$ has gap $\Delta_A$.  }
\label{Junction}
\end{figure}
We assume a situation such as the one displayed in Fig. (\ref{Junction}). We treat the superconductor $S$ in the tunnel region. We shall later see that beyond the instability point two gaps appear in $S$ as depicted in the figure. For simplicity $S_A$ may be regarded as being in equilibrium.

In addition, we shall need the assumption of local equilibrium, namely that the system thermalizes on a much faster time scale than it is forced out of equilibrium. The non-equilibrium forcing is associated with the injection of imbalance into the sample. The injection process is balanced out in steady state by processes that relax imbalance, which involve excitations  scattering to a region close to the gap. We shall assume throughout that $|T-T_c| \ll T_c$. In this regime  $\Delta \ll T_c$  and only a fraction $\frac{\Delta}{T}$ of the processes involve excitations near the gap. This results in  a typical branch imbalance relaxation rate which is given by  $\tau_Q^{-1} = \tau_\epsilon^{-1} \frac{\Delta}{T}$, satisfying $\tau_Q \gg \tau_\epsilon$. Local equilibrium establishes on a time scale $\tau_\epsilon$ if the non-equilibrium forcing is small enough such that it operates on a time scale $\tau_Q$ along with imbalance relaxation processes.

When these conditions are satisfied namely when $\tau_\Delta \ll \tau_\epsilon \ll \tau_Q$, one may use the following distribution function, $n$, for the occupation number of excitations, written here through the linear combination $\rho=2n-1$ for convenience:
\begin{align}\label{simpleDistribution}
\rho(\xi)\equiv 2n-1 = \tanh \left( \frac{\epsilon(\xi) + \Phi q(\xi) }{2 T_c}\right) + \delta \rho,
\end{align}
where  $\epsilon(\xi) = \sqrt{(\xi-\xi_1)^2 + \Delta^2}, $ is the BCS spectrum of excitations and $q(\xi)  = \frac{\xi-\xi_1}{\epsilon(\xi)}$, is the charge of an excitation, and $\delta \rho$ is the next to leading order correction in $\frac{\Delta}{T}$. The distribution function (\ref{simpleDistribution}) describes a situation where the condensate is at chemical potential $\xi_1$, while the quasi-particles have a chemical potential $-\Phi$ above it. The possibility to sustain the two entities at different potentials exists due to the fact that on the thermalization time scale $\tau_\epsilon$, transitions from the condensate to the quasi-particle population are suppressed, since the latter only happen on a time scale $\tau_Q$.

Quasi-neutrality dictates $\xi_1 = \Phi$. To find the relation between $\xi_1$ and $\Delta$ one may use the self-consistency condition,  $\int \frac{\rho}{\epsilon} d\xi = \frac{2}{g},$ for $g$ the interaction coupling constant (see (\ref{Hamiltonian}).  This gives \cite{galaiko} \cite{spivak}:
\begin{align} \label{Supression}
\Delta^2 + 2 \Phi^2 + T_c^2 \int \frac{\delta\rho}{\epsilon} d\xi = c T_c \delta T,
\end{align}
where $c= \frac{8 \pi^2}{7 \zeta(3)}$, and $\delta T = T_c - T$. This is a corrected version of (\ref{simplesupression}) The charge imbalance $\Phi$ may be given an order of magnitude estimate by balancing the injection rate to the relaxation rate  of imbalance, $\tau_Q^{-1}$. This is given by $\Phi = \frac{b}{2 \sqrt{2}} I \tau_\epsilon T_c^2 \Delta^{-1}$, for some constant $b$ of order unity. Here $I$ is the injection intensity, namely, the inverse time it takes for an excitation to enter the superconductor.  The contribution of the integral in (\ref{Supression}) can also be estimated based on a expansion of the kinetic equation by order of $\Delta \over T$ \cite{PethickSmith}\cite{spivak}. The result is  $a T_c^2 I \tau_\epsilon$, for some constant $a$ of order unity. This constant may be either positive or negative depending on whether the injection process tends to suppress or enhance \cite{ChiClake} superconductivity, respectively. We shall assume a situation where $a$ is negative \cite{spivak}, namely enhancement \footnote{Enhancement only serves to provide a situation where steady state is possible beyond the stability point. In the opposite case $a>0$ an instability of the kinetic negates such a steady state solution \cite{spivak}. This instability is unrelated to the the instabilities of  (\ref{GorkovsEquaitons}), which this Letter addresses.}.

\begin{figure}
\includegraphics[width=3.0in]{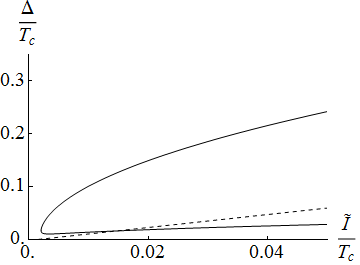}
\caption{The two branched dependence of $\Delta$ on the injection rate $\tilde{I}$, both in units of $T_c$,  for $c \delta T= -0.002$, $a=-1.23$ and $b=0.276$ . The state is not stable under the dashed line.}
\label{EnhancemtCurves}
\end{figure}
Putting the above estimates into (\ref{Supression}) one obtains a relation between the gap $\Delta$ and the injection rate:
\begin{align}\label{DeltafunctionofI}
2\Delta^2 = \Delta_p^2 \pm \sqrt{\Delta_p^4- ( b\tilde{I} )^2 },\quad \Delta_p^2 =   c T_c \delta T - a \tilde{I}  ,
\end{align}
where $\tilde{I} = I T_c^2 \tau_\epsilon$. For a given $\tilde{I}$ there exist two different $\Delta$ corresponding to the $+$ and $-$ signs in (\ref{DeltafunctionofI}).  Using a stability analysis of the semiclassical equations (\ref{GorkovsEquaitons}) it was shown in \cite{spivak} that the superconductor becomes unstable to cooper pairing for $\Delta< \frac{\Delta_p^2}{T_c}$, this occurs when the injection rate is above some critical value, and only for  the solutions in (\ref{DeltafunctionofI}) corresponding to the minus sign.  It was also shown in \cite{spivak} that beyond this point the order parameter starts oscillating on the time scale $\tau_\Delta$. This invalidates the distribution function (\ref{simpleDistribution}) along with  the kinetic approach based on the BCS expressions for the spectrum $\epsilon(\xi)$ and excitation charge $q(\xi)$. The BCS approach is no longer valid as it is based on the assumption that the order parameter is time independent on a time scale $\tau_\Delta$.  We shall argue in the sequel that (\ref{simpleDistribution}) is in fact valid, albeit with a new spectrum of excitations, consisting of two gaps, and a suitable $q(\xi)$, both computed making use of Richardson's exact solution of the problem.

\section{Richardson's solution} Richardson \cite{Richardson} had solved exactly the pairing Hamiltonian (\ref{Hamiltonian})
The first stage for solving the Hamiltonian is to note that singly occupied levels do not participate in the dynamics and are blocked to the pairs. One may simply drop the singly occupied levels from the spectrum and solve (\ref{Hamiltonian}) for the case where levels are either occupied or unoccupied with pairs.
\begin{figure}[h]
\includegraphics[width=3in]{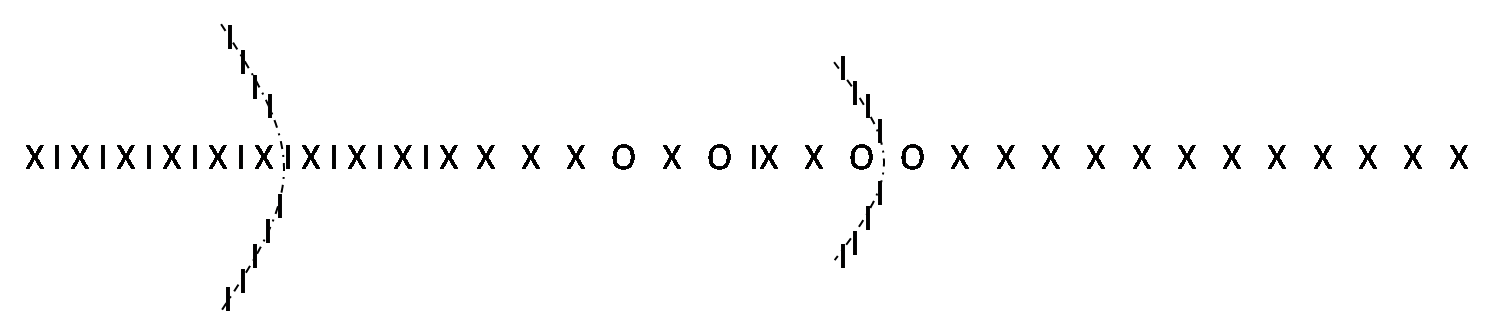}
\caption{Two dimensional plane occupied with charges. $X$ and $O$  denote unblocked an blocked levels respectively. $I$'s denote the $E$'s.\label{RichardsonFig}}
\end{figure}

We take the Hamiltonian (\ref{Hamiltonian}), where $j$ runs over $2M$ levels around the Fermi energy. Suppose that there are $N+2M$ particles in the system and $L$ levels are blocked. Richardson's solution for the eigenstates for the reduced spectrum is the following:
\begin{align}
\prod_{\alpha=1}^{\frac{N+2M-L}{2}} b^\dagger_\alpha |0 \>,\quad b^\dagger_\alpha = \sum_i \frac{1}{E_\alpha - \xi_i} c^\dagger_{i, \uparrow} c^\dagger_{i, \downarrow}
\end{align}
Where the $E$'s are {\it complex} parameters which must satisfy the following non-linear equations:
\begin{align}\label{Richardson}
\frac{2}{g} = - \sum_j \frac{1}{E_\nu - \epsilon_j } + \sum_{\mu\neq \nu} \frac{2}{ E_\nu-E_\mu},
\end{align}
which  can be viewed as the condition of vanishing electric field (equilibrium) for each of the charges $E_\nu$, in two-dimensions. Each  $E$ carries a charge $+1$, while at each location of $\xi_i$ a charge $-\frac{1}{2}$ is present. In addition, there is a background field, $-\frac{1}{g}$. An equilibrium point where  $E$ can be placed can always be found between any two unblocked $\xi$'s. In the continuum limit, where the level spacing is the smallest energy scale, this defines a charge density, $\lambda(\xi)$, on the real axis, which is composed of the contribution of both the $E$'s and the unblocked $\xi$'s. Assuming a constant density of state $\rho_0$ for the $\xi_i$'s, we may also define $\rho(\xi) = 2 \rho_0^{-1} \lambda(\xi)$. In addition to the $E$'s on the real axis some of the $E$'s arrange themselves in the complex plane. In the continuum limit the distribution of the complex $E$'s is given by line densities on arcs in the complex plane \cite{Sierra}\cite{Gaudin}. The distribution of the $E$'s may be termed as the `Richardson spectrum` (See Fig. (\ref{RichardsonFig})). The position and shape of the arcs is determined by electrostatic equilibrium. $\rho(\xi)$ may be understood as the generalization of the occupation number of excitations in the BCS approach. Indeed, To effect an excitation one must change the density $\rho(\xi)$.

For a given $\rho(\xi)$, and assuming $m$ arcs in the complex plane harboring the complex $E$'s,  a continuum limit solution for the charge distribution is given in the following form \cite{Sierra} \cite{Gaudin}:
\begin{align}
h(\xi) = \int \frac{R(\xi) \rho(\xi')}{R(\xi')(\xi'-\xi)} d\xi', \quad R(\xi) = \sqrt{\prod_{i=1}^m (\xi-\xi_i)^2 + \Delta_i^2},  \nonumber
\end{align}
The jump discontinuity of the electric field are caused by the line density of the charges. The jump discontinuities of $h$ consists of a jump discontinuity on the real axis, and jump discontinuities on the branch cuts of $R(\xi)$ which are drawn as curved lines stretching from $\xi_i +i \Delta_i$ to $\xi_i - i \Delta_i$. The values of $\xi_i$ and $\Delta_i$ are constrained by self-consistency conditions:
\begin{align}\label{selfconsistencies}
\int \frac{\rho \xi^l }{R(\xi)} = \frac{2}{g}\delta_{l,m-1} , \quad l\leq m-1.
\end{align}
Note that for $m=1$ we get the usual self-consistency condition, where $\rho(\xi)$ is to be identified with $2n(\xi)-1$.

The energy of the Richardson state is given by $\sum_\nu E_\nu + \sum_j \xi_{i_j}$, where $\xi_{i_j}$ are the blocked states. As $h(\xi)$ encodes the density of $E$'s, it is possible to compute the energy and the number of particles (here $R(\xi) = R(\xi)$):
\begin{align}
E=   \int  \frac{\left[ \xi R(\xi)  \right]_+  \rho(\xi) }{R(\xi)} d\xi
,\quad N=   \int  \frac{\left[ R(\xi) \right]_+  \rho(\xi) }{R(\xi)} d\xi
\end{align}
where  $[\dots]_+$ denotes taking the non-negative (polynomial in $\xi$) part of the Laurent expansion around infinity.

\section{Quasi-particles around instability point}
To describe the Cooper instability in the framework of Richardson's approach. Consider a normal metal state placed below $T_c$. Since $\Delta=0$ no arcs exist and all the $E$'s are on the real axis, distributed such that the total charge density corresponds to Fermi distribution. The instability manifests itself as a vanishing of  $h(\xi)$ at two complex conjugated points. Since the field vanishes, these points may be occupied by $E$'s. Indeed, collisions will cause these to start populating until a full arc is formed, describing the equilibrium superconducting state, as described in Fig. (\ref{Sketch}).

By inserting (\ref{simpleDistribution}) into the the expression for $h(\xi)$, one sees that points of vanishing field appear for  $\Delta < \frac{\Delta_p^2}{T_c}$ (this conclusion was reached by semiclassical means in  \cite{spivak}). The resolution of the instability is achieved when an arc appears near the Fermi energy, in addition, that is, to the arc at $\xi_1$.

We now study the state with two arcs, $m=2$. By Hellmann-Feynman,  $q(\xi_i) = \frac{\delta E}{\delta \xi_i}$, assuming that the distance between the two arcs is much larger than their size, one gets: $q(\xi) = \frac{(\xi-\xi_1)(\xi-\xi_2)}{R(\xi)}$. Note that this form of $q(\xi)$ dictates the following excitation representation: Both well below $\xi_1$ and well above $\xi_2$, $q(\xi)=1$ and excitations are electron-like. In the region between $\xi_1$ and $\xi_2$ but well away from both gaps, $q(\xi)=-1$, and excitations are hole-like. This choice does not agree with the standard choice of the excitation representation. However,  the choice of excitation charge far away from the gaps is a matter of representation.

Consider now transitions due to collisions between states characterized by two arcs. To compute the transition rates, one must compute the matrix element of the phonon interaction between two different Richardson states. We are interested in solving the kinetic equation only to first order in $\Delta \over T$, so one may assume that one of the excitations involved in the transition, has energy of order $T$, and thus has the same character as in the normal metal. Under such an assumption, the transition of this excitation to a level near the gap is dictated by the one particle density matrix of the target level. The one particle density matrix consistent with $q(\xi)$ computed above, can be easily found. From its form, it is easy to see that,  the transition rates may be defined by coherence factors just in the BCS case, and these are given by $u^2(\xi) = \frac{1}{2} (1-q)$, $v^2(\xi) = \frac{1}{2}(1+q)$.

The other ingredient in a kinetic approach is the energy conservation delta function which features in the Fermi golden rule. To study energy conservation we make use of the following picture valid close to $T_c$ due to Pethick and Smith \cite{PethickSmith}, which states that a change in the number of particles may be divided into a contribution of the change of the normal charge, $ \delta q_n =\int q \delta \rho d\xi $  and superconducting charge, $ \delta q_s =\int \rho \delta q d\xi $,   as follows $\delta N =  \delta q_s + \delta q_n$. Starting from the expression for the energy of the level, we now write down the variation with respect to a change in the density $\rho$ of the combination $W=E - \nu N$. Consider the energy of an excitation next to $\xi_i$:
\begin{align}
\delta W = \left( \frac{ \sqrt{(\xi-\xi_i)^2 +\Delta_i^2}}{(-)^i} + (\xi_i-\nu) q(\xi) \right) \delta \rho + (\xi_i-\nu) \delta q_s, \nonumber
\end{align}
the  $(-)^i$ arises from the unusual choice of the excitation representation discussed above (we assume $\xi_1<\xi_2<0$).  The energy of an excitation, neglecting the charge transfer to the condensate, is written as $ \epsilon_\nu (\xi) = \left. \frac{\delta W}{\delta \rho} \right|_{\delta q_s =0}$.

The kinetic equation may now be written and solved taking into account the energy change due to charge transferred to the condensate ($\delta q_s$). It turns out however, that these contribute to a lower order. As a result the distribution function turns out to be $\rho(\xi) = \tanh\left(\frac{ \epsilon_{\nu=0}(\xi)}{2 T_c} \right) +\delta \rho$.  $\delta \rho$ is found by balancing the injection rate with the relaxation provided by the linearized collision operator. The linearization is performed around the normal-metal state, so the relaxation times are independent of $\Delta$, which yields that $\delta \rho$ in (\ref{simpleDistribution})  depends only on $I$ and temperature.

\begin{figure}
\includegraphics[width=3.0in]{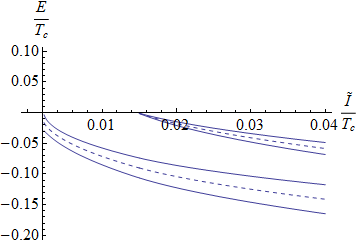}
\caption{The dependence of the two gaps on the injection $\tilde{I}$. The dashed lines are $\xi_1$ and $\xi_2$ ($\xi_2$ lies above $\xi_1$). The solid lines are (from bottom to top), $\xi_1-\Delta_1$, $\xi_1+\Delta_1$, $\xi_2-\Delta_2$, $\xi_2+\Delta_2$, and delineate the gaps. The approximations leading to this solution become poorer as the gaps approach one another.  }
\label{ResultsFigure}
\end{figure}
To obtain $\xi_1$ and $\xi_2$ we separate the contribution of the super-current  from the region close to $\xi_1$ in energy and the region close to $\xi_2$.  We demand that the charge carried away as supercurrent associated with a particular region is equal to the amount of charge transferred to the condensate by collision into that region. The superconductor velocity $v_s$ is the same regardless of the region considered, while the contribution to the density of superconducting pairs from region $i$ is  proportional to $\Delta_i^2$. Collisions  to region $i$ transfer charge to the condensate  proportional to $\xi_i \Delta_i$ . As a result, one obtains $\xi_i  = -\frac{b}{2 \sqrt{2}} \tilde{I} \frac{\Delta_i}{\Delta_1^2 +\Delta_2^2}$. Plugging these estimates into the self consistency equations (\ref{selfconsistencies}) one obtains equations relating the gaps to the current intensity:
\begin{align}
\Delta_i^2 + 2 \xi_i^2   = c T_c \delta T - a \tilde{I} -T_c \Delta_{\bar{i}} \left|\frac{ \xi_{\bar{i}}}{\xi_{\bar{i}}-\xi_i}\right|,
\end{align}
where $\bar{1}=2$ and $\bar{2}=1$. A sketch of the results for the same parameters for which Fig. (\ref{EnhancemtCurves}) is drawn is given in Fig. (\ref{ResultsFigure}).

\section{Conclusion}
We have seen that non-equilibrium effects may excite a mode by which a superconductor develops another gap. The possibility of having two gaps  is most easily revealed by considering the multi-arc Richardson states. These gaps will manifest themselves in the same type of experiments that reveal the BCS gap. A probe for the density of states (such as another tunnel junction) must be coupled to the injection region. In situations where the current exceeds the instability point, a multi-gapped structure is predicted.  The multi-gapped state also manifests itself in oscillations of the order parameters. Experiments involving the Josephson junction coupled to the injection region will be sensitive to these oscillations. The form of  oscillations can be found  using the quantum to classical correspondence between Richardson's states and the semiclassical solutions found in \cite{kuznetsov}.  Applying this relation to an exact description of the oscillatory behavior  will be the subject of future work, however without further analysis one can conclude that, due to the fact that the system describes two condensates at energies separated by a distance $\delta \xi$, the oscillations will predominately be at a frequency of $\frac{2 \delta \xi}{\hbar}$.  We have developed the theory in the simplest case where the two condensates are well separated at a temperature near $T_c$. It is interesting to extend the theory beyond this simple application. This may be done as a kinetic theory is possible due to the fact that the transition rates between Richardson states and the energy of any such state is known or  feasibly computable \cite{Calabrese}. The complications of solving the kinetic theory may be overcome by using a numerical approach.

In the specific situation discussed in this paper two spectral gaps appeared. The theory, however, extends to cases where more than two gaps exist. Roughly speaking, additional gaps appear at points, $\xi_0$, where $n(\xi_0)=1/2$ and pairing correlations are small enough as not to suppress the appearance of a new gap. Experimental setups can potentially be created that would deform the distribution function so strongly as to create such points (the distribution function must be made to pass through $1/2$ and the other gaps must be suppressed). Given the distribution function, a quantitative treatment can be given using the generalized self-consistency conditions, (\ref{selfconsistencies}). 

\section{Acknowledgement}
I would like to acknowledge many helpful discussions with B. Spivak, A. Nahum, M. Moshe, G. Gorohovski, F. Rashed, B. Laikhtman, O. Agam and D. Orgad. This research was funded by ISF  grant  206/07.

\bibliography{mybib}{}
\bibliographystyle{apsrev4-1}

\end{document}